\documentclass{llncs}
\usepackage{llncsdoc}
\usepackage{graphicx}
\usepackage{subfigure}
\usepackage{makeidx}
\usepackage{hyperref}
\usepackage{amssymb}
\begin{document}

\newcounter{save}\setcounter{save}{\value{section}}
{\def\addtocontents#1#2{}%
\def\addcontentsline#1#2#3{}%
\def\markboth#1#2{}%
\title{A New P2N Approach to Software Development Under the Clustering}
\author{Gang Liao \and Lei Liu \and Lian Luo}

\institute{Computer Science and Engineering Department,
 Sichuan University Jinjiang College, 620860
 Pengshan, China\\
 GreenHat1016@Gmail.com,cys19900611@gmail.com,mr.l172586418@gmail.com}

\maketitle
%
\begin{abstract}
In this computer era of rapid development, software development can be seen everywhere, but a lot of softwares are dead in modern development of software. Just as \textit{The Mythical Man-Month} said, it exists a problem in the software development, and the problem is interflow. A lock of interflow can be said great calamity. Clustering is a environment to breed new life. In this thesis, we elaborate how P2N can be used to thinking, planning, developing, collaborating, releasing. And the approach that make your team and organization more perfect.
\end{abstract}

\section{Introduction}
Point to Node (P2N) is a use in a development team, he can be solved in the development of the design, coding, testing, and maintenance work. The main use of the cluster of ideas, can be easily software evaluation, team communication, and reconstruction.

Software engineering in the 1960s, the software is far behind the development of the hardware. The use of the software developers who use, no universal value has become the bottleneck in the development of computer systems. So that time there have been software workshops, that is, software development, in order to meet the needs of users, to improve software development.

Clustering is a computer system for link incompact computers, to complete highly closely cooperation computational work.The cluster is a group of service work together to provide scalability and availability service platform more than a single service entity. Classification or clustering, as it is more often referred as, is a data mining activities that aim to differentiate groups inside a given set of objects, being considered the most important. Clustering is to complete the same job across multiple computers to achieve higher efficiency and the two PCs or more PCs; the process is exactly the same. If one crashes, another can work.

Cluster as early as long ago put forward, and now, more for high performance computing. Such as, High-availability (HA) clusters, Load balancing clusters, High-performance clusters (HPC), Grid computing and so on. In 2006, a paper by Istv¨¢n Gergely Czibula introduction use a clustering approach to improve system design. At the same time, Clustering-Based support for software architecture restructuring was published by Niels Streekmen in 2011. Point to Node (P2N) is used software development, make full use of the integrity of the set and close communication. Frederick P. Brooks, Jr. has proposed the interflow is an essential part of the software, and little communication or do not communication in a large proportion of the failure of the software development.

All existing integrated Development Approach offers support for automatic application of various refactoring. In this thesis we are focusing on  developing a technique that would help developer to identify the appropriate refactering.Discussed in this paper, the ideal way to P2N (point to node) is caused by a web-based P2P development team. Clusters are generally used for network communications, shared resources. In the development, communication exchange is very important, just that we use this characteristic of the cluster communication, build a development framework, with P2N this approach, the development problems have been avoided.

Our approach takes an existent software and maintain using clustering. in order to obtain a better design, suggesting the needed refactoring remains the decision of the software development.\cite{1}

\section{Software development}
Successful software systems are long-living systems\cite{7}. The modernization of existing software systems is an important topic in software engineering research and practice. The determination of the software for the whole project came with the crucial, also is the positioning accuracy. Otherwise, its likely development software has passed\cite{8}.
\begin{figure}
\centering
\includegraphics[width=0.7\textwidth]{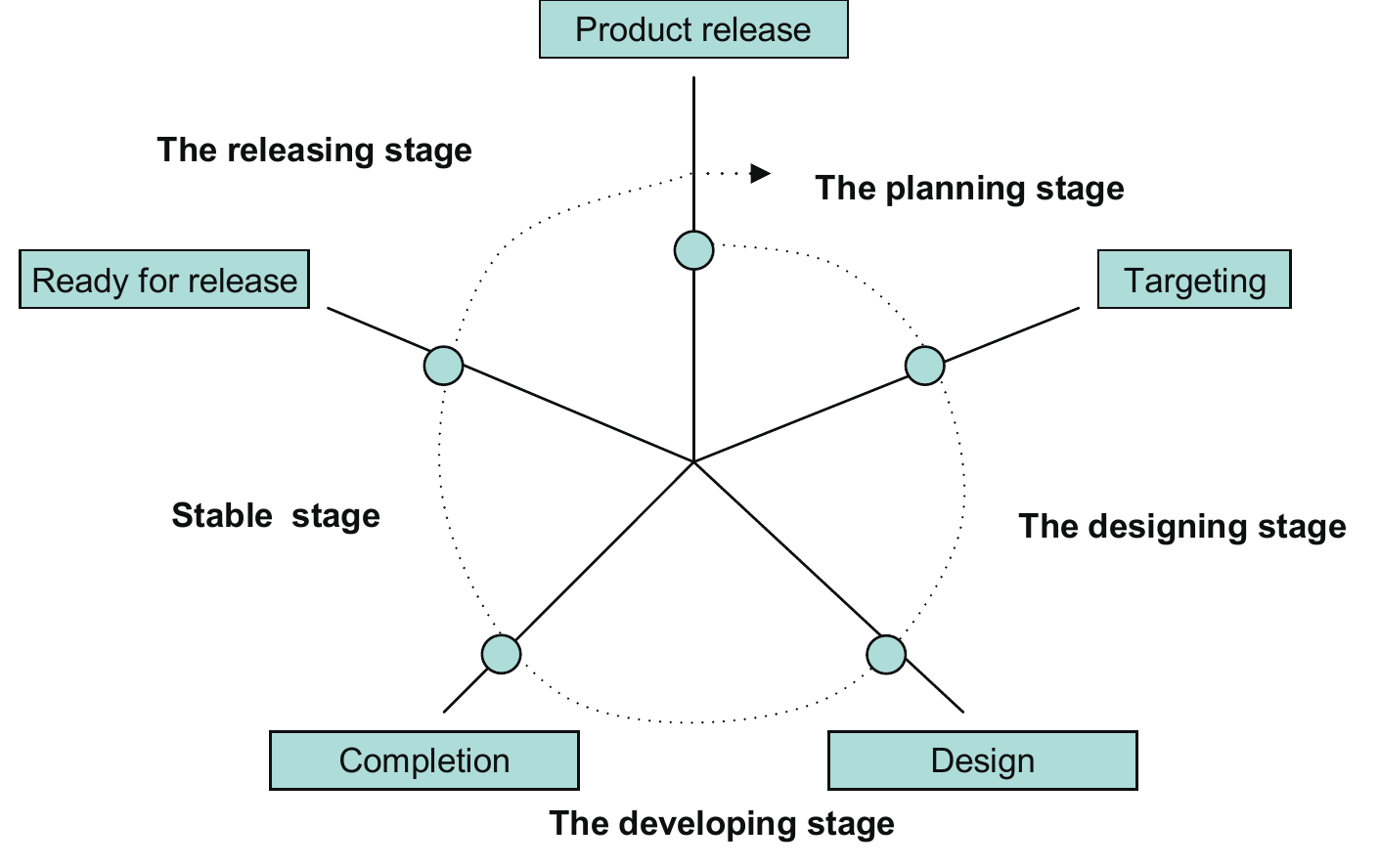}
\caption{Software development process.}
\label{figc1}
\end{figure}
We can see clearly that will pass of software development target, design, code, test, and maintenance, etc several stages\cite{9}.
 \begin{center}
\begin{tabular}{|c|c|}
\hline
\textbf{Software task} & \textbf{Rate}\\
\hline
Plan & 1/3\\
\hline
Code & 1/6\\
\hline
Test & 1/4\\
\hline
Maintenance & 1/4\\
\hline
\end{tabular}
\end{center}
 with the improvement of the requirement, testing in proportion of the increase graduality. Although the development personnel in the planning and do not wish to plan too much time on the testing, testing occupies half of the time entire software development. Communication is important in software development, developer could know course of task using P2N (Point to Node). Consciousness of team in the development of nowhere not be reflected. Now, Object-Oriented(OO) is used by most software development, in order to better service to user.P2N (Point to Node) is a approach and based on the thought of the development.

In the feasibility study, we use the less the cost to accomplish more, and more reliable task, our initial idea can be achieved. To clarify the target problem, the logical model of the export system, starting from the logical model, using various algorithms to achieve various functions. This is truly the time of a project from requirements to test, among which is the methods and algorithms developers to identify an effective way to achieve the function of each part and to make each function you want to do possible improvement.

\section{Point of Software}
Software is made of objects, a object is made of classes, a class is made of methods and attributes. The essentials of this definition are plan, in the mind, and later execution. Thus, a design is a created object and completed methods. Software is seen 3D diagram.
\begin{figure}
\centering
\includegraphics[width=0.7\textwidth]{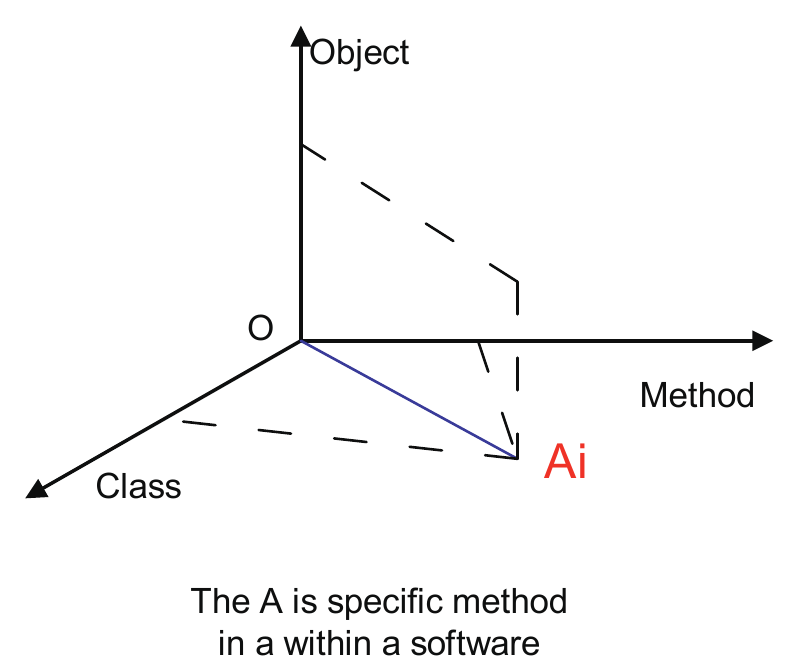}
\caption{Abstract Model.}
\label{figc1}
\end{figure}
Fig.\ref{figc1} shows software between 3D model of the internal relations.
\begin{math}
S = \{O_{1},O_{2} ,\dots{},O_{n} \}
\end{math}
be the set of objects to be clustering. Using the Vector-space model, each object is measured with respect to a set of \textit{l} initial attributes,
\begin{math}
Ar = \{Ar_{1},Ar_{2},\dots{},Ar_{n}\}
\end{math}
,and is therefore described by a A-dimensional vector
\begin{math}
O_{i} = \{O_{i1},O_{i2},\dots{},O_{i\textit{l}}\},O_{ih}\in R,1 \leq i \leq n,1\leq h \leq \textit{l};
\end{math}
 Usually, the attributes associated to objects are standardized in order to ensure an equal ``path'' to all of them(the path equal weight ) Using Euclidain distance
 \begin{math}
 d: O\times O\rightarrow R.
 \end{math}
 The distance between two objects expresses the dissimilarity between them Consequently, the similarity between two object \begin{math} O_{a} \end{math} and \begin{math}O_{b} \end{math} is defined as:
 \begin{displaymath}
 sim(O_{a},O_{b}) = \frac{1}{d(O_{a},O_{b})}
 \end{displaymath}
 The computation of the similarity between a newly formed cluster and the existing clusters in each step of hierarchical clustering is often not newly computed based on the similarity of the single nodes, but using an updating rule. Describe the four basic strategies for this computation.

\begin{itemize}
\item[] \begin{math}
 Single Linkage  \qquad  sim(C_{i},C_{jk}) = max(Sim(C_{i},C_{j}),Sim(C_{i},C_{k}));\end{math}
\item[]
\begin{math}Complete Linkage  \qquad Sim(C_{i},C_{jk}) = Min(Sim(C_{i},C_{j})),Sim(C_{i},C_{k});\end{math}
\item[]
\begin{math}Weighted Average Linage \qquad sim(C_{i},C_{jk}) = \frac{sim(C_{i},C_{j}),\frac{1}{2 *sim(C_{i},C_{k})}}{2}\end{math}
\item[]
\begin{math}
Unweighted Average Linage \qquad sim(C_{i},C_{jk}) = \frac{sim(C_{i},C_{j}) * size(C_{j}) + sim(C_{i},C_{k}) * size(C_{k})}{size(C_{j},C_{k})}\end{math}\cite{2}
\end{itemize}
The \begin{math}A_{i}\end{math} shows a method in the specific class of a particular object. In other words
\begin{math}
A_{i} = O_{j}\bigcap C_{k}\bigcap M_{h} = \{<x,y,z>\in O\times C\times M \mid O\bigcup C\bigcup M \in R\}
\end{math}
Every point is discrete in the 3D diagram.
\begin{figure}
\centering
\includegraphics[width=0.7\textwidth]{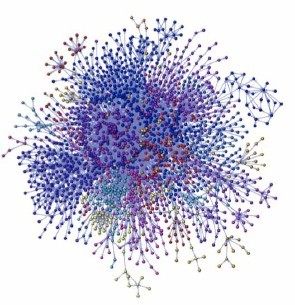}
\caption{Software internal contact.}
\label{figc2}
\end{figure}
Fig.\ref{figc2} is imitated to software, every single point is looked at a method and the line between two points show the relation between them. Different branch means different class and different color expresses different object. Every branch is different subsystem. A software graph (3D network)is a digraph O = (V, L) that consist of the set V of classes and the set of relationships
\begin{math}L = P\bigcup S\end{math}.
 There are two type of relationships: a membership relationship
\begin{math}P = \{<V_{i},V_{j}> \mid V_{i}\in V_{j}\}\end{math}, read ``
\begin{math}V_{i}\end{math}
 has part
 \begin{math}V_{j}\end{math}'';
and a reflexive and transitive relationship \
\begin{math}S = \{<V_{i}, V_{j}> \mid V_{i}\in V_{j}\}\end{math}, read ``
\begin{math}V_{i}\end{math}
is a subclass of
\begin{math}V_{j}\end{math}''
\cite{3}. Software graphs show an important information space of Object-Oriented software system\cite{10}. Software graphs are highly heterogeneous net works, where a very few classes participate in many relationships and the majority of classes have one or two relationships.Software graphs show an important information space of Object-Oriented software system. Software graphs are highly heterogeneous net works, where a very few classes participate in many relationships and the majority of classes have one or two relationships\cite{4}.
And software internal is fairly complex. We can get a local to analysis in Fig.\ref{figc2}.We can clearly see the relation in Fig.\ref{figc3}. a1 is a Node and Node is the next part of the content to discuss. We will be called a1 root node, it shows a subsystem.
\begin{figure}
\centering
\includegraphics[width=0.7\textwidth]{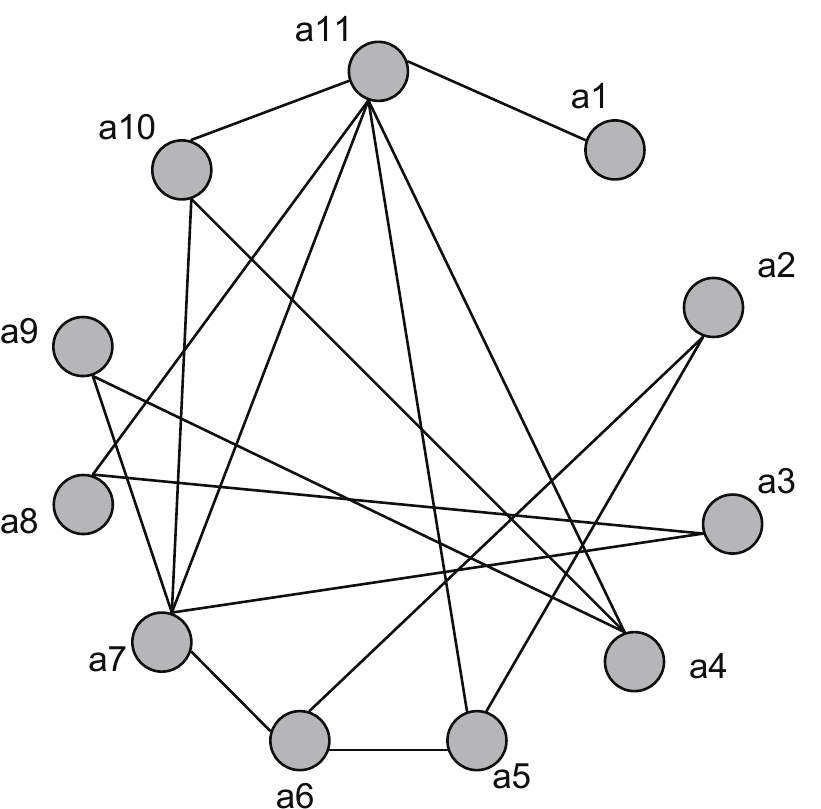}
\caption{Software partial diagram.}
\label{figc3}
\end{figure}
We will be this subsystem remember N1, and
\begin{math}
N_{1} = \{m_{1},m_{2},\dots{},m_{11}\}\end{math},software is made of \begin{math}N_{n}\end{math}, expressed as: \begin{math}S = \{N_{1}, N_{2},\dots{}, N_{n}\}.\end{math}

\section{Point to Node (P2N)}
Clustering techniques have already been applied for program restructuring. In \cite{5} a clustering based approach for program restructuring\cite{1}. A cluster can be simulated to become a software test platform, where each node corresponds to the different parts. Like the link between class and class in java and C$++$ development software, object properties and methods. This can be reflected in the cluster. We now discuss the P2N is a small branch, the purpose is to improve the degree of coupling between the internal, and to effectively implement the reconstruction and maintenance. They said point-to-node, instead of the traditional peer-to-peer, but the principle is almost the same, the role is different. 

In this section, we present an overview of the Point to Node (P2N) which is used for developing approach of software system. The traditional point to point is that each user plays than the color of the server, not just download a single task, but also shoulder the task of uploading resources. Based on P2P, we think of P2N P2N is the idea of clusters in the LAN (local area network) will be able to get a good application, each developer can easily understand the progress of the entire team in their development at the same time, being in this can be adjusted at their own pace, previously used in the form of the document to develop communication, can now directly use P2N the ways and means to carry out the exchange between projects, to achieve the true sense: without leaving and know everything. We have already mentioned above, many software developers are also mentioned, communication is an integral part of a development, the failure of many projects are out of communication problems, the groups appear in the "generation gap".P2N this problem is relatively solved easily resolve such drawbacks.Point to Node (P2N) is point of software to node of clustering. Clustering usually used to data mining, the purpose is to distinguish between a given a set of object.
\begin{figure}
\centering
\includegraphics[width=0.7\textwidth]{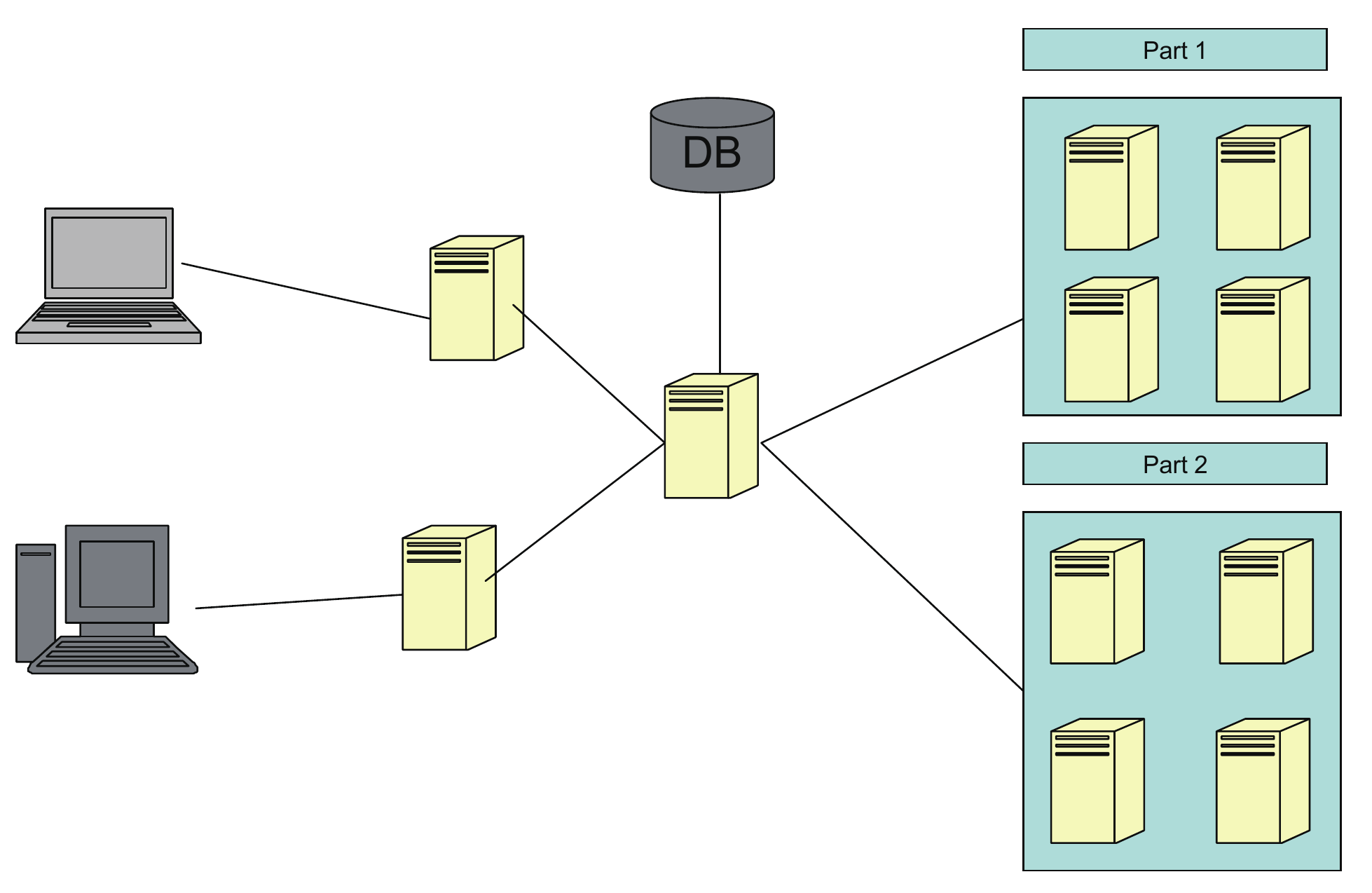}
\caption{Schematic of the clustering.}
\label{figc4}
\end{figure}

Fig.\ref{figc4}shows a practical cluster, right part is cluster system. Every little part is made of many PCs. We borrow diagram to present that Node of clustering. For big companies, each programmer only develops a class or a method. Code correctness, completeness, reusability demanding. Unit test regards as Node, There are specific node contain a class or the realization of the method.In order to measure software each part of ``saturation''\cite{11}. Programmers write code to submit the unit node. It has its own special test and mechanisms to detect the presentation of the integrity of the code. In this stage of the unit test, that is, the code submitted by the sub-stage. During this period, we can better correction code, and a series of code refactoring. Software development and building are the same, the earlier the better. Using P2N approach is detailing the problem, so that the trouble has become more aware and the developer is easy to find.

We want to call other¡¯s code but would like to call others, its considerable trouble. Using of P2N (Point to Node) this way, clearly on the child node display on the platform. The convenient the ¡°code¡± of communication between the developer. Sine P2N approach focus on providing better under standing of the development, they tend to present refactoring code. As the coed evolves, some approaches focus on the co-evolution of the refactoring code.\cite{6}
\begin{center}
\begin{tabular}{|c|c|c|c|}
\hline
\textbf{} & \textbf{Corrective } & \textbf{Adaptability} & \textbf{Perfection of } \\
\textbf{} & \textbf{maintenance} &\textbf{ maintenance} & \textbf{maintenance}\\
\hline
\hline
Intelligibility & $\surd$ & &\\
\hline
Testability & $\surd$ & & \\
\hline
Modifiability & $\surd$ & $\surd$ &\\
\hline
Reliability & $\surd$ & &\\
\hline
Portability & & $\surd$  &\\
\hline
Usability & & $\surd$ & $\surd$ \\
\hline
Efficiency & & & $\surd$ \\
\hline

\end{tabular}
\end{center}
P2N of the features is the ability to clearly reflect the progress of development, "communication" in the development in each branch node set of tests, the corresponding test in accordance with the requirements of the project\cite{12}. Software development are blind, small-scale projects are more prominent, they do not have a good set of development guidelines, and developed software from the user's initial is the most a team effort wasted.

\section{Application method of P2N}
 Message Passing Interface (MPI),this is a common build cluster technology.The massage passing interface effort began in the summer of 1991 when a small group of researchers started discussions at a mountain retreat in Austria. MPI is a language-independent communications protocol used to program parallel computer.Both point-to-point and collective communication are supported. MPI is a message-passing application programmer interface, together with protocol and semantic specification for how is features must behave in any implementation.
 
 It is technology to build a cluster. Most MPI implementations consist of a specific set of routines directly callable from C,$C++$,Fortran and any language able to interface with such libraries,including $C\#$,Java, or Python\cite{13}.We will develop a platform to build the MPI cluster, a compiler more to see.Framework as a cone, the layers of depth, layers of testing, layers to modify.Solve the same problems or more complex issues in a short period of time; can reduce inputs; execute multiple instructions at the same time.By increasing the computing power within a local node, making it the so-called "super nodes", not only improves the speed of development of the whole system, computing capacity, and can improve system modularity and scalability.The ultimate goal is the development of testing and refactoring the problem is mapped to the cluster, this mapping is achieved through the different levels of abstraction mapping.P2N development design, cluster of the problem needs to be transformed into a specific pattern of development of suitable cluster model, in order to reach this goal, the first is the problem of clustering algorithms must be able to problem internal cluster features fully reflected, otherwise the cluster solution methods can not be made use of these cluster characteristics, so that quickly become impossible.Second, each pivot point in the development and cluster development model try to find a consistent way, the premise of the problem on a cluster solution.
 
 The two most cluster programming model is a data parallel and message-passing, data parallel programming mode, high-level programming is relatively simple, but it is only used for data parallel problems\cite{14}; programming level message passing programming model is relatively low, but the message-passing programming model can have a broader range of applications.Message-passing between the implementation of part of the cluster by passing messages to water the flowers, coordinating the pace control the execution.Messaging for the programmer to provide a more flexible means of control and expression, the number of data parallelism is difficult to achieve, are implemented using message passing.

Messages for the programming on one hand provide flexibility, on the other hand, it will also perform various crisscross of complicated exchange of information between and coordination, control task transmitted to the programming schedule, and, to some extent, is the increased programmer burden, but for the system laid a solid fortress. So in P2N programming technique, we adopted is message transfer programming model.

\section{Discussion and conclusion}
In this paper we have asked the question of how use the clustering to improve speed of software development. Sine successful system are doomed to continually evolve and grow, P2N approach should support co-evolution mechanisms to keep all recovered views. In this paper we presented a approach for software development. We adopt a directed clustering to provide a unified environment  for the software development to test and maintain.From a platform, a method to determine the integrity and artistry of the things to do.

Now ,our approach is to build on the basis of the cluster, we can use parallel technology to software testing, software architecture reconstruction based on the cloud, software engineering come from the real meaning of growth and development. Finally,our vision is that software testing.Software development will be toward a better trend toward a new vision.There is intelligent inspection, testing the integrity of, multi-faceted.The emergence of cloud computer has to a higher area, real hidden information, improve information security.Next, we would cloud further explore and analyze software problems now appear to identify a better way to remove the software bottleneck.

These results are discouraging,suggesting that,for large system, such as we used as a basis for our software development, and for the approach we chosen,more work needs to be done before these clustering is ready to be widely adopted.However, it may be that such approach is useful in less stringent environment.Our hope is that our results spur to further efforts both to create automated clustering and to subject these approach to solve the problems of software development such as we reported in this thesis.

\end{document}